\documentclass{vldb}
\usepackage{graphicx}
\usepackage{balance}  
\usepackage{subcaption}
\usepackage{tabularx}
\usepackage{array}
\usepackage{booktabs}
\usepackage{rotating}
\usepackage{cite}
\usepackage{url}
\usepackage{enumitem}
\usepackage{float}
\usepackage{placeins}
\usepackage{xspace}
\usepackage{hyperref}
\usepackage{afterpage}
\afterpage{\clearpage}

\newcommand{\system}{SmartQueue\xspace}

\usepackage{xcolor}

\begin{document}


\title{Buffer Pool Aware Query Scheduling \\via Deep Reinforcement Learning \\(Extended Abstract)}

\numberofauthors{1} 

\author{Chi Zhang$^{1}$, Ryan Marcus$^{2}$, Anat Kleiman$^{1}$, Olga Papaemmanouil$^{1}$ \\
\affaddr{$^1$Brandeis University \quad $^2$MIT CSAIL and Intel Labs} \\
\affaddr{$^1$\{chizhang, akleiman, opapaemm\}@brandeis.edu, $^{2}$ryanmarcus@csail.mit.edu}
 }

\maketitle

\begin{sloppypar}
\begin{abstract}
 In this extended abstract, we propose a new technique for query scheduling with the explicit goal of reducing disk reads and thus implicitly increasing query performance. We introduce \system, a learned scheduler that leverages overlapping data reads among incoming queries and learns a scheduling strategy that improves cache hits. \system relies on deep reinforcement learning to produce workload-specific scheduling strategies that focus on long-term performance benefits while being adaptive to previously-unseen data access patterns. We present results from a proof-of-concept prototype, demonstrating that learned schedulers can offer significant performance improvements over hand-crafted scheduling heuristics. Ultimately, we make the case that this is a promising research direction in the intersection of machine learning and databases.  

\end{abstract}
\end{sloppypar}
\begin{sloppypar}
\section{Introduction}

Query scheduling, the problem of deciding which of a set of queued queries to execute next, is an important and challenging task in modern database systems. Query scheduling can have a significant impact on query performance and resource utilization while it may need to account for a wide number of considerations, such as cached data sets, available resources (e.g., memory), per-query performance goals, query prioritization, or inter-query dependencies (e.g., correlated data access patterns). 

In this work, we attempt to address the query scheduling problem by leveraging overlapping data access requests. Smart query scheduling policies can take advantage of  such overlaps, allowing queries to share cached data, whereas naive scheduling policies may induce unnecessary disk reads. For example, consider three queries $q_1, q_2, q_3$ which need to read disk blocks $(b_1, b_2)$, $(b_4, b_5)$, and $(b_2, b_3)$ respectively. If the DBMS's buffer pool (i.e., the component of the database engine that caches data blocks) can only cache two blocks at once, executing the queries in the order of $[q_1, q_2, q_3]$ will result in reading 6 blocks from disk. However, if the queries are executing in the order $[q_1, q_3, q_2]$, then only 5 blocks will be read from disk, as $q_3$ will use the cached $b_2$. Since buffer pool hits can be orders of magnitude faster than cache misses, such savings could be substantial.

In reality, designing a query scheduler that is aware of the current buffer pool is a complex task. First, the exact data block read set of a query is not known ahead of time, and is dependent on data and query plan parameters (e.g., index lookups). Second, a smart scheduler must balance short-term rewards (e.g., executing a query that will take advantage of the current buffer state) against long-term strategy (e.g., selecting queries that keep the most important blocks cached). One could imagine many simple heuristics, such as greedily selecting the next query with the highest expected buffer usage, to solve this problem. However, a hand-designed policy to handle the complexity of the entire problem, including different buffer sizes, shifting query workloads, heterogeneous data types (e.g., index files vs base relations), and balancing short-term gains against long-term strategy is much more difficult to conceive.

Here, we showcase a prototype of \system, a deep reinforcement learning (DRL) system that automatically learns to maximize buffer hits in an adaptive fashion. Given a set of queued queries, \system combines a simple representation of the database's buffer state, the expected reads of queries, and deep Q-learning model to order queued queries in a way that garners long-term increases in buffer hits.
\system is fully learned, and requires minimal tuning. \system custom-tailors itself to the user's queries and database, and learns policies that are significantly better than naive or simple heuristics. In terms of integrating \system into an existing DBMS, our prototype only requires access to the execution plan for each incoming query (to assess likely reads) and the current state of the DBMS buffer pool (i.e., its cached data blocks).

We present our system model and formalized our learning task in Section~\ref{s:model}. We present preliminary experimental results  from a proof-of-concept prototype implementation in Section~\ref{sec:experiments}, related work in Section~\ref{s:related}, and in Section~\ref{sec:conclusion} we highlight directions for future work.



\section{The SmartQueue Model}\label{s:model}


 \system is a learned query scheduler that  automatically learns how to order the execution of queries to minimize disk access requests. The core of \system includes a deep reinforcement learning (DRL) agent~\cite{deep_rl} that learns a query scheduling policy through continuous interactions with its environment, i.e., the database and the incoming queries. This DRL agent is not a static model, instead it \emph{continuously} learns from its past  scheduling decisions and \emph{adapts} to new  data access and caching patterns. Furthermore, as we discuss below, using a DRL model allows us to define a reward function and scheduling policy  that captures long-term benefits vs short-term gains in disk access.  
 
\begin{figure}
  \centering
  \includegraphics[width=0.45\textwidth]{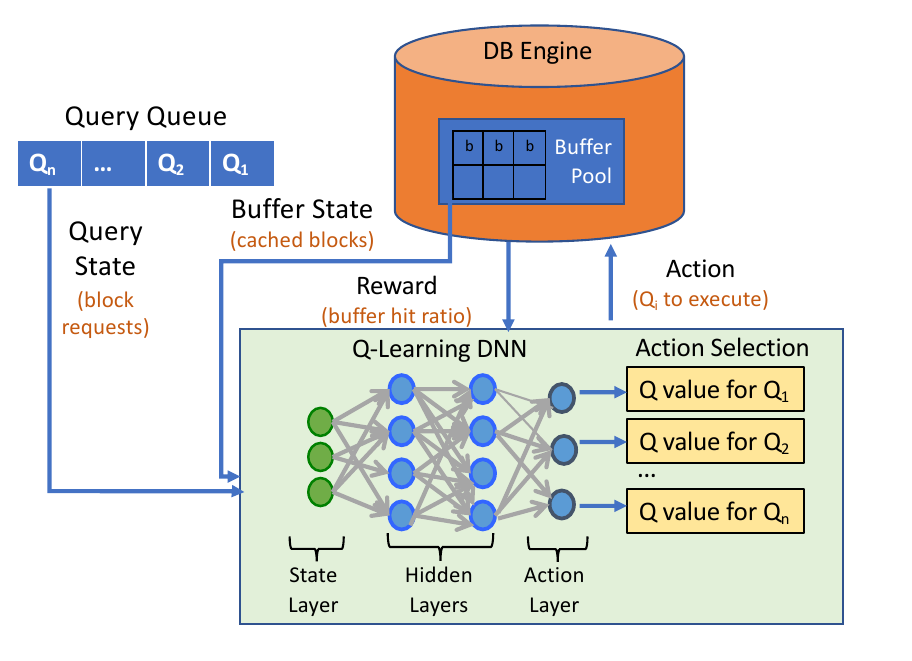}
    \vspace{-2mm}
  \caption{\system's system model}
  \label{fig:system_model}
  \vspace{-5mm}
\end{figure}

Our system model is depicted in Figure~\ref{fig:system_model}. Incoming user queries are placed into an execution queue and \system decides their order of execution. For each query execution, the database collects the required \emph{data blocks} of each input base relation, where a data block is the smallest data unit used by the database engine. Data blocks requests are first resolved by the buffer pool. Blocks found in the buffer (\emph{buffer hits}) are returned for processing while the rest of the blocks (\emph{buffer misses}) are read from disk and placed into the buffer pool (after possible block evictions). Higher buffer hit rates (and hence lower disk access rates) can enormously impact query execution times but require strategic query scheduling, as execution ordering affects the data blocks cached in the buffer pool. 

One tempting solution to address this challenge could involve a greedy scheduler which executes the query that will re-use the maximum number of cached data blocks. While this simple approach would yield short term benefits,  it ignores the long-term impact of each choice. Specifically, while the next query for execution will maximally utilize the buffer pool contents, it will also  lead to newly cached data blocks, which will affect future queries. A greedy approach fails to identify whether these new cached blocks could be of any benefit to the unscheduled yet queries.

\system addresses this problem by training a deep reinforcement learning agent to make scheduling decisions that maximize long term benefits. Specifically, it uses a model that simultaneously estimates and tries to improve a weighted average between short-term buffer hits and the long-term impact of query scheduling choices.  In the next paragraphs,  we discuss the details of our approach: (a)  the input features vector that capture data access requests (\emph{Query Bitmap}) and buffer state (\emph{Buffer Bitmap}), and (b) the formalized  DRL task. 

\paragraph*{Buffer Bitmap} One input to the DRL model is the state of the buffer pool, namely which blocks are currently cached in memory. Buffer state $B$  is represented by a bitmap where rows represent base relations and columns represent data blocks. The $(i,j)$ entry is  set to 1 if the $j$-th block of relation $i$ is cached in the buffer pool and is set to zero otherwise.  Since the number of blocks of any given relation can be very high and different for each relation, each row vector $F_i$ is downsized by calculating a simple moving average over the number of its blocks entries. Specifically $D_i$ is the downsized row of a relation $i$ and  $F_i$ is the full size row, we have:
\begin{equation}
 B_{ij} = \lfloor|F_i|/|D_i|\rfloor \times \sum_{k=j \times \lfloor|D_i|/|F_i|\rfloor}^{(j+1) \times \lfloor|D_i|/|F_i|\rfloor} F_{ik}
\end{equation}
\paragraph*{Query Vector} The second input to the DLR model is the data block requests of each  query in the queue. Specifically, given a query $q$, we generate a vector  that indicates the data blocks to be accessed by $q$ for each base relation in the database. To implement this,  \system collects the query plan of $q$, and approximates the probability of each table's data block being accessed. Our approach handle requests of index file and base relations similarly, as both type of blocks will be cached into the buffer pool. The query vector is downsized in the same was as the buffer bitmap. 

Full table scans for a base relation $i$ indicate that all data blocks of the given relation will be accessed, and  therefore each cell of the $i$-th row vector has the value of 1. For indexed table scans, we calculate the number of tuples to be accessed based on the selectivity of the index scan. If the index scan is feeding a loop-based operator (i.e., nested loop join) the selectivity is adapted accordingly to account for any iterations over the relation. We assume the relation is uniformly stored across data blocks and therefore, if $x\%$ tuples of a base relation are to be selected from an indexed operation, we set the access probability of each data block of the relation to $x\%$. Similarly, we assume that the indexed operation reads $x\%$ of the index's blocks. We note that much more sophisticated probabilistic models could be used, but for this preliminary work we use this simple approximation.

\paragraph*{Deep Q-Learning} \system uses deep Q-learning~\cite{deep_learning} in order to decide which query to execute next.  As with any deep reinforcement learning system, \system is an agent that operates over a set of states $S$ (buffer pool states) and a set of actions $A$ per state (candidate queries to executed next). \system models the problem of query scheduling as a Markov Decision Process (MDP)~\cite{rl_book}: by picking one query from the queue to execute, the agent transitions from the current to a new buffer pool state (i.e., data blocks cached). Executing a new query on the current buffer state, provide the agent with a reward. In our case, the reward of an action is the buffer hit ratio of the executed query calculated as $\frac{\mbox{\textit{buffer hits}} }{\mbox{\textit{total block requests}}}$.

The goal of the agent is to learn a \emph{scheduling policy} that maximizes its total reward. This is a continuousus learning process: as more queries arrive and the agent makes more scheduling decisions, it collects more information (i.e., context of the decision and its reward) and adapts its policy accordingly.
The scheduling policy is expressed as a function $Q(S_t,A_t)$, that outputs a \emph{Q-value}  for taking an action $A_t$ (i.e., a query to execute next) on a buffer state $S_t$. Given a state $S_t$ and a action $A_t$, the Q-value $Q(S_t, A_t)$ is calculated by adding the maximum reward attainable from future buffer states to the reward for achieving its current buffer state, effectively influencing the current scheduling decision by the potential future reward. This potential reward is a weighted sum of the expected buffer hit ratios of all future scheduling decisions starting from the current buffer state. Formally, after each action $A_t$ on a state $S_t$ the agent learns a new policy $Q^{new}(S_t, A_t)$ defined as:
\begin{equation}
 Q(S_t, A_t)+\alpha[R_{t} + \gamma \max_{a}{(Q(S_{t+1},a)-Q(S_t, A_t))}]
\end{equation}
The parameter $\gamma$ is the discount factor which weighs the contribution of short-term vs. long-term rewards. Adjusting the value of $\gamma$ will diminish (e.g., favor choosing queries that will make use of the current buffer state) or increase (e.g., favor choosing queries that will allow long-term increased usage of the buffer) the contribution of future rewards. The parameter  $\alpha$ is the learning rate or step size. This simply determines to what extent newly acquired information overrides old information: a low learning rate implies that new information should be treated skeptically, and may be appropriate when a workload is mostly stable but contains some outliers. A high learning rate implies that new information is more fully trusted, and may be appropriate when query workloads smoothly change over time. Since the above is a recursive equation, it starts with making arbitrary assumptions for all $Q$-values (and hence arbitrary initial scheduling decisions). However, as more experience is collected through the execution of incoming queries, the network likely converges to the optimal policy~\cite{dqn}.

\section{Preliminary Results} \label{sec:experiments}

Here, we present preliminary experiments demonstrating that \system
can generate query ordering that increase the buffer hit ratio and improve query execution times compared with alternative non-learned schedulers. 

\paragraph*{Experimental Setup} Our experimental study used workloads  generated using the 99 query templates of the TPC-DS benchmark~\cite{tpcds}. We deployed a database with a size of 49GB on single node server with 4 cores, 32GB of RAM. For our experiments, we generated $1,000$ random query instances out of these 99 templates and placed them in a random order in the execution queue.  The benchmark includes 165 tables and indexes, and the number of blocks for each of these ranged between $100$ and $130,0000$. However, after downsizing  both the query vector and buffer state bitmaps, our representation vectors have a size of {$165 \times 1,000$}, including index tables. 
 We run our experiments on PostgreSQL~\cite{url-postgres} with a shared buffer pool size of 2GB.\footnote{We configured PostgreSQL to bypass the OS filesystem cache. In future work, multiple levels of caching should be considered.} For each query, we collect its query plan without executing the query by using the \texttt{EXPLAIN} command.

\system uses  a fully-connected neural network. Our DRL agent was implemented with Keras\cite{keras} and uses 2 hidden layers with 128 nerons each. We also use an adaptive learning rate optimization algorithm (Adam~\cite{adam}) and our loss function is the mean squared error.  

In our study, we compare SmartQueue with two alternative scheduling approaches. \emph{First-Come-First-Served (FCFS)} simply executes queries in the order they appear in the queue. \emph{Greedy} employs a simple heuristic to identify the query with the best expected hit ratio given the current contents of the buffer pool. Specifically, for each queued query it calculates the dot product of the buffer state bitmap with the data requests bitmap, estimating essentially the probability of buffer hits for each data block request. We then order all queries based on the sum of these probabilities over all blocks and execute the query with the highest sum value. Following the execution, the new buffer state is calculated and the heuristic is applied again until the queue is empty. This greedy approach focuses on short-terms buffer hits improvements.  

\begin{figure*}[t]
  \centering
  \begin{subfigure}{0.45\textwidth}
    \includegraphics[width=\textwidth]{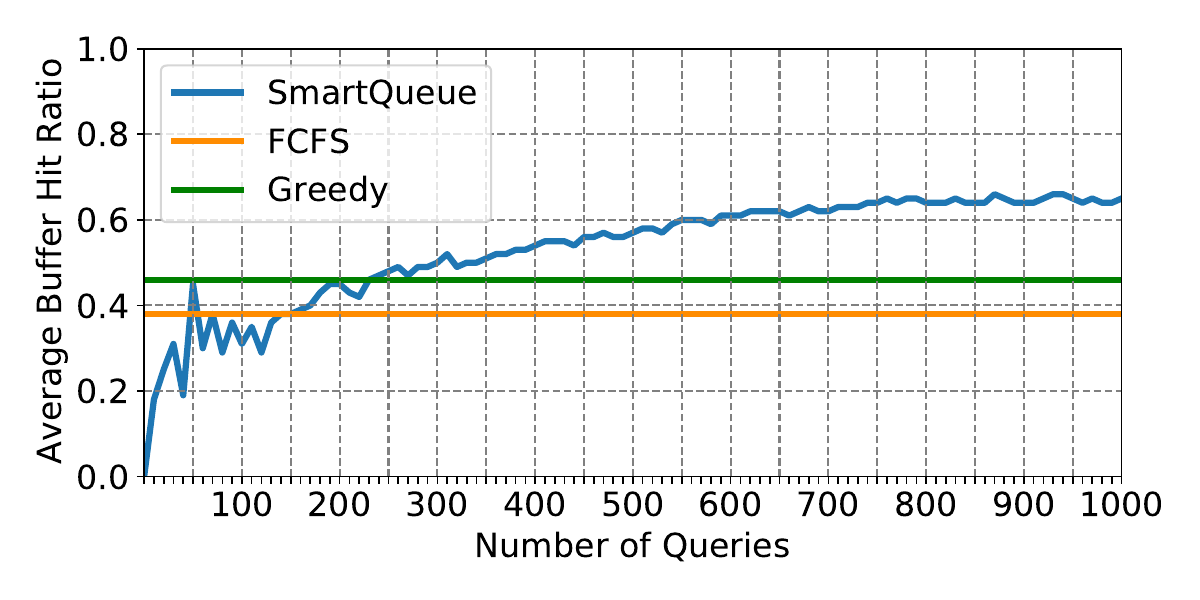}
    \caption{Average buffer hit ratio}
    \label{fig:e2_buffer}
  \end{subfigure}
      \begin{subfigure}{0.45\textwidth}
    \includegraphics[width=\textwidth]{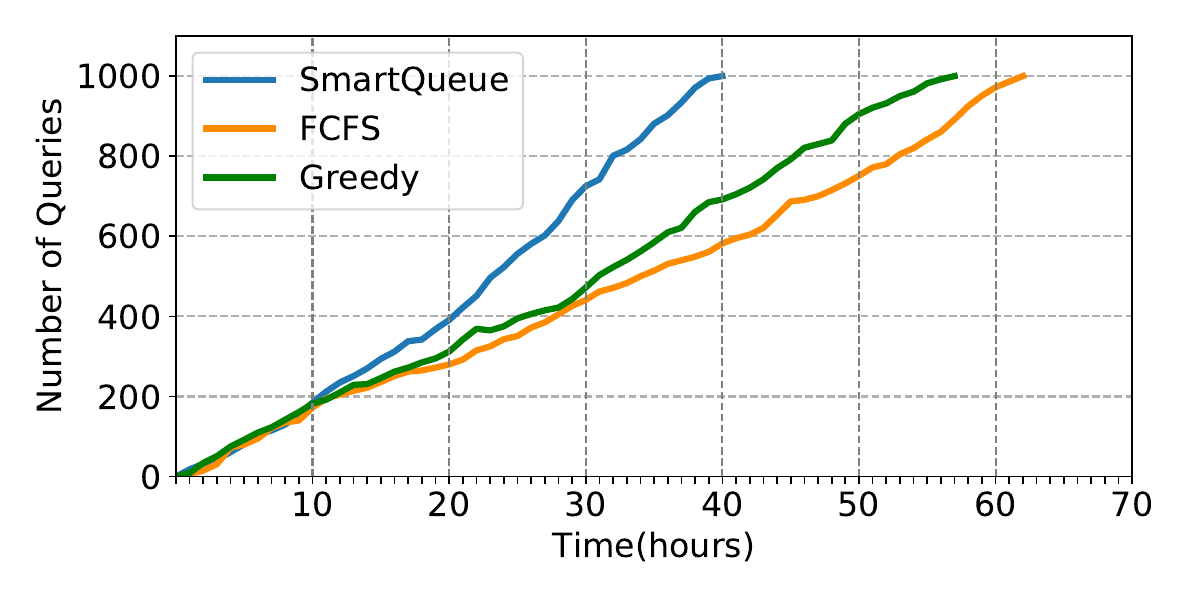}
    \caption{Query execution time}
    \label{fig:e2_latency}
  \end{subfigure}
  \caption{\system's effectiveness (buffer hit ratio and query completion rate) with increasing training sets. }
  \label{fig:no_split}
\end{figure*}

\paragraph*{\bf Effectiveness}

First, we demonstrate that \system can improve its effectiveness as it collects more experience.  In this set of experiments, we placed all $1,000$ queries in the queue and we start scheduling them using \system.  In the beginning our agent will make arbitrary scheduling decisions, but as it schedules more queries, \system collects more experience from its past actions and starts improving its policy. To demonstrate that, we evaluated the learned model at different stages of its training. Figure~\ref{fig:e2_buffer} and Figure~\ref{fig:e2_latency} shows how the model performs as we increase the number of training queries. In Figure~\ref{fig:e2_buffer}, we measure the average buffer hit ratio  when scheduling our $1,000$ queries and we compare it with the buffer hit ratio of FCFS and Greedy (which is not affected by the number of training queries). We observe that the DRL agent is able to improve the buffer hit ratio as it schedules more queries. It outperforms the buffer hit of the other two heuristics eventually converging into a ration that is  65\% higher than  FCFS and $35\%$ higher than Greedy.

In addition, Figure \ref{fig:e2_latency} shows the number of executed queries over time. The results demonstrate that DRL-guided scheduling of \system allows our approach to execute the workload of $1,000$ queries around $42\%$ faster than Greedy and $55\%$ faster than FCFS. This indicates that \system can effectively capture the relationship between buffer pool state and data access patterns, and leverage that to better utilize the buffer pool and improve its query scheduling decisions.

\paragraph*{\bf Adaptability to new queries} 

\begin{figure*}[t]
  \centering
  \begin{subfigure}{0.45\textwidth}
    \includegraphics[width=\textwidth]{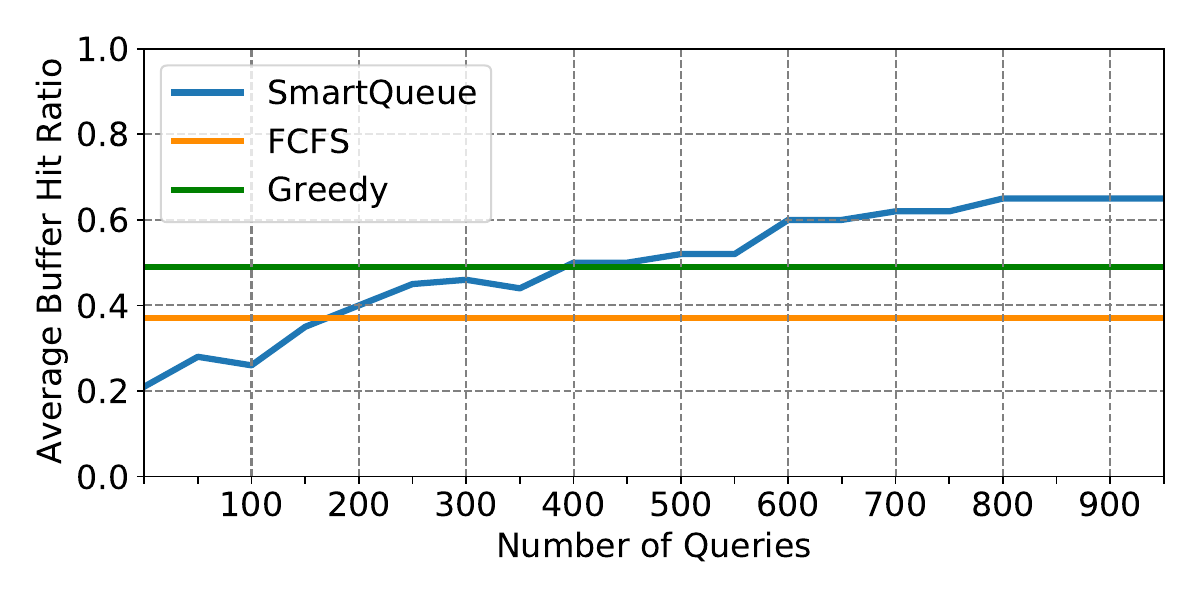}
    \caption{Average buffer hit ratio }
    \label{fig:e1_buffer}
  \end{subfigure}
      \begin{subfigure}{0.45\textwidth}
    \includegraphics[width=\textwidth]{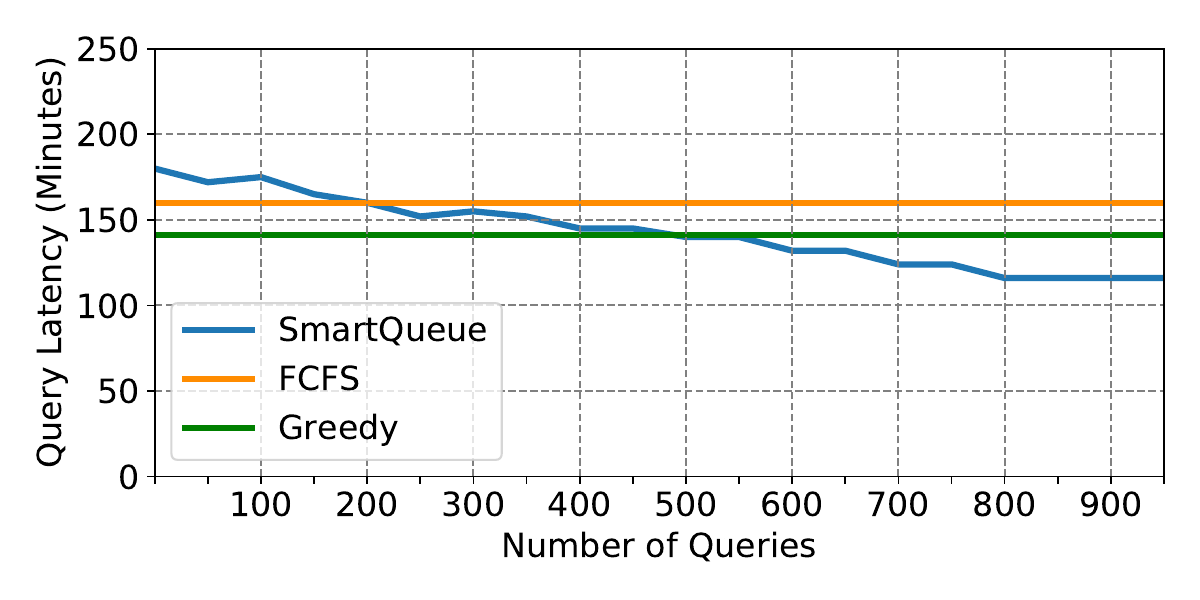}
    \caption{Query execution time}
    \label{fig:e1_latency}
  \end{subfigure}
  \caption{Buffer hit ratio and latency improvement on unseen query templates and increasing training queries. }
  \label{fig:train_test_split}
\end{figure*}

Next we studies \system's ability to adapt to unseen queries. For these experiments, we trained \system by first scheduling $950$ random queries out of 79 TPC-DS templates. We then test the model over 50 random queries out 20 unseen before TPC-DS templates. Figure~\ref{fig:e1_buffer} demonstrates how  average buffer hit ratio of the testing queries is affected as \system collects experience increases from scheduling more training queries. The graph shows that the average buffer hit ratio of the testing queries is increased from 0.2 (when the \system is untrained) to 0.64 (when \system has schedule all 950 queries). Furthermore, \system outperforms FCFS and Greedy after having scheduled less than $500$ queries. 

Finally, Figure ~\ref{fig:e1_latency}, shows that the query latency of our testing queries keeps decreasing (and eventually outperforms FCFS and Greedy) as \system is trained on more queries. Our approach enables unseen queries to be eventually executed $11\%$ faster than FCFS and $22\%$ than Greedy. These results indicate that  query scheduling policy can adapt to new query templates leading to significant performance and resource sharing improvements. 

\paragraph*{\bf Overhead}
We also measured the training and inference time. Our proof-of-concept prototype needed 240 mins to incorporate 950 queries in our agent (so in average the training overhead is $3.95$ mins per query). This time does not include the execution time of the query. This training overhead can potentially be optimized by offloading it into another thread, introducing early stopping, or re-using previous network weights to get a good "starting point." There is no training overhead for FCFS and Greedy.  The inference time of \system is {3.12}seconds  while the inference time for Greedy is 2.52 seconds and {0.0012} seconds for FCFS.


\section{Related work} \label{s:related}

Prior work on query scheduling have focused on query parallelism~\cite{q-cop}, elastic cloud databases~\cite{azar, cost_wait, leitner, wisedb-cidr, pmax, sqlvm, nashdb}, meeting SLAs~\cite{icbs,sla-tree,wisedb-vldb,slos,perfenforce_demo,sloorchestrator,activesla,smartsla}, or cluster scheduling~\cite{decima,opennebula,step}. In terms of buffer pools and caching, most prior work has focused on smart cache management~\cite{cache-augment,cache-tables} (i.e., assuming the query order is fixed and choose which blocks to evict or replace), or on (memory) cache-aware algorithms~\cite{mem-cache}. Here, we take a flipped approach, in which we assume the buffer management policy is fixed and the query order may be modified (e.g., batch processing). 
More broadly, work on learned indexes follows recent trends in integrating machine learning components into systems~\cite{pillars}, especially database systems. Machine learning techniques have also been applied to query optimization~\cite{neo, skinnerdb, qo_state_rep}, cardinality estimation~\cite{deep_card_est2, naru, plan_loss}, cost modeling~\cite{learn_cost}, data integration~\cite{termite, deep_entity}, tuning~\cite{ml_tuning}, and security~\cite{sql_embed}.


\section{Conclusion and future work}
\label{sec:conclusion}

We have presented \system, a deep reinforcement learning  query scheduler that seeks to maximize buffer hit rates in database management systems. While simple, \system was able to provide substantial improvements over naive and simple heuristics, suggesting that cache-aware deep learning powered query schedulers are a promising research direction. \system is only an early prototype, and in the future we plan to conduct a full experimental study of \system. In general, we believe the following areas of future work are promising.

\vspace{1mm} \noindent \textbf{Neural network architecture.}  While effective in our initial experiments, a fully connected neural network is likely not the correct inductive bias~\cite{inductive_bias_ml} for this problem. A fully connected neural network is not likely to innately carry much useful information for query scheduling~\cite{inductive_bias_rl}, nor is there much of an intuitive connection between a fully-connected architecture and the query scheduling problem~\cite{inductive_bias_dl}. The first layer of our network learns one linear combination per neuron of the entire input. These linear combinations would have to be extremely sparse to learn features like "the query reads this block, which is cached." Other network architectures -- like locally connected neural networks~\cite{locally-connected} -- may provide significant benefit. 

\vspace{1mm} \noindent \textbf{SLAs.} Improving raw workload latency is helpful, but often applications have much more complex performance requirements (e.g., some queries are more important than others). Integrating query priorities and customizable Service Level Agreements (SLAs) into \system by modifying the reward signal could result in an buffer-aware and SLA-compliant scheduler. 

\vspace{1mm} \noindent \textbf{Query optimization.} Different query plans may perform differently with different buffer states. Integrating \system into the query optimizer -- so that query plans can be selected to maximize buffer usage -- may provide significant performance gains. 

\vspace{1mm} \noindent \textbf{Buffer management.} \system only considers query ordering, and assumes that the buffer management policy is opaque. A larger system could consider both query ordering and buffer management, choosing to evict or hold buffered blocks based on future queries. Such a system could represent an end-to-end query scheduling and buffer management policy.


\bibliographystyle{abbrv}
\bibliography{ryan-cites-short}
\balance

\end{sloppypar}
\end{document}